\documentclass{article}
\usepackage{graphicx}
\usepackage{amsmath}
\def\bq{\begin{equation}}
\def\eq{\end{equation}}
\def\bqa{\begin{eqnarray}}
\def\eqa{\end{eqnarray}}
\def\roughly#1{\mathrel{\raise.3ex
\hbox{$#1$\kern-.75em\lower1ex\hbox{$\sim$}}}}
\def\lsim{\roughly<}

\def\gsim{\roughly>}

\def\be{\begin{equation}} 
\def\ee{\end{equation}} 
\def\bq{\begin{equation}} 
\def\eq{\end{equation}} 
\def\bqa{\begin{eqnarray}} 
\def\eqa{\end{eqnarray}} 
\def\lsim{\roughly<}
\def\gsim{\roughly>}
\def\nle{\mathrel{\vcenter
     {\hbox{$<$}\nointerlineskip\hbox{$\sim$}}}}
\def\nge{\mathrel{\vcenter
     {\hbox{$>$}\nointerlineskip\hbox{$\sim$}}}}

\begin{document}

%%%                               Preprint number
%\begin{flushright}
%KEK-CP-???\\
%CUCP-10-??? \\
%KU-PH-???
%\end{flushright}

%%%                               Title
\begin{center}
{\large \bf One-loop effects of MSSM particles \\
 in $e^-e^+\to Zh$ and $e^-e^+\to \nu\bar\nu h$ at the ILC}

\end{center}
\vspace{0.2 cm}
\begin{center}

Yusaku Kouda$^{1,*}$, Tadashi Kon$^1$,  Yoshimasa Kurihara$^2$, Tadashi Ishikawa$^2$, \\ 
Masato Jimbo$^3$, Kiyoshi Kato$^4$ and Masaaki Kuroda$^5$\\
\vspace{4 mm}
$~^{1}~$ Seikei University, Musashino, Tokyo 180-8633, Japan\\
$~^{*}~$ E-mail: {dd146101@cc.seikei.ac.jp} \\
$~^{2}~$ KEK, Tsukuba, Ibaraki 305-0801, Japan\\
$~^{3}~$ Chiba University of Commerce, Ichikawa, Chiba 272-8512, Japan\\
$~^{4}~$ Kogakuin University, Shinjuku, Tokyo 163-8677, Japan\\
$~^{5}~$ Meiji Gakuin University, Yokohama, Kanagawa 244-8539, Japan\\
\end{center}

\noindent{\bf Abstract}\\
\ The 1-loop effects of the MSSM at the ILC are investigated through numerical analysis.  We studied the higgs production processes $e^-e^+\rightarrow Zh$ and $e^-e^+\rightarrow \nu\bar{\nu}h$ at the ILC. It is found that the magnitude of the MSSM contribution through the 1-loop effects is sizable enough to be detected. In the study, three sets of the MSSM parameters are proposed,
which are consistent with the observed higgs mass, the muon $g$-$2$,
the dark matter  abundance and the decay branching ratios of $B$ mesons. 
In the $e^-e^+\rightarrow Zh$ process, the 1-loop effects of the MSSM are visible and the distinction of the parameter sets is partially possible. For the study of  $e^-e^+\rightarrow \nu\bar{\nu}h$, we used the equivalent $\it W$-boson approximation in the evaluation of the 1-loop cross section. While the 1-loop effect of the MSSM is visible, the distinction of the parameter sets might not be possible in this process under the value of realistic luminosity at the ILC. 

%Within the framework of the MSSM, we investigate the 1-loop effects
%of  SUSY  particles in the higgs production processes at the ILC. Three sets of the SUSY parameters are proposed 
%which are consistent with the observed higgs mass, the muon $g$-$2$,
%the dark matter  abundance and the decay branching ratios of $B$ mesons. We investigated the 1-loop effects of SUSY in $e^-e^+ \to Zh$ and $e^-e^+\to\nu \bar \nu h$  at the ILC. In the latter analysis we used effective $W$ bosson approximation.
%We discuss on the possibility of discovering signals consistent with SUSY as well as of distinguishing  the proposed
%sets of SUSY parameters experimentally. 

\vspace{0.5cm}

\section{Introduction}
　The standard model (SM) is completed by the discovery of the last piece, the higgs particle. However, it is argued that it is not the final theory of the fundamental particles.
For example, the SM includes a lot of free parameters. When one calculates the mass of higgs, the values of these parameters are intentionally selected to cancel the quantum correction. This cancellation is as precise as up to around 17 order. Some researchers regard it as  ``unnatural".  \\ 
　The supersymmetric (SUSY) model \cite{martin} is considered as one of the promising  candidates for the theory beyond the standard model. 
In this theory, each particle in the SM has its supersymmetric partner, or, a sparticle.
The quadratic divergence in the calculation of quantum correction to the higgs mass is canceled by other contributions from the sparticles, so that the fine tuning problem disappears.
The search of sparticles is the important subject of the present and future collider experiments to prove the SUSY.
In spite of hard efforts in the large hadron collider (LHC) experiments, slightest signature of their existence has not been obtained.
For example, the scalar top particle (stop) seems not to exist under $\mathcal O$(1)TeV mass region \cite{atlastop,cmsstop}.
Though sparticles are so heavy that it is difficult to be produced directly, the indirect signature would be expected at the international linear collider (ILC).

%For this issue, the SUSY provide us the clue. In this theory, the quadratic divergences in the calculation of quantum correction of higgs mass are canceled out with the contribution of new particles. These particles are called ``sparticles" which is predicted as a pair for all of SM particle. The searches of sparticles are important subjects of the present and future  collider experiments. 
%In spite of very precious efforts of participants of the large hadron collider (LHC),
%even slight signature of their existence have not been obtained. For example, it have been cleared that the scalar top particle(stop) does not exist under $\mathcal O$(1TeV).
% If sparticles is so heavy that it is difficult to be produced directly or the sensitivity is too weak to be discovered, the signature would be expected at international linear collider(ILC). 

%On the other hand, the only weakly interacted sparticles must be under $\mathcal O$(0.5TeV), if we naively consider the low energy experiments such as muon $g$-$2$. Moreover, in order to be consistent with the dark matter thermal relic density, the mass spectra of MSSM is quite limited. 
%The first object of this study is finding the parameter sets which meets these experimental bounds and cosmological observation under these very special situation.
%The second object is confirming the possibility of the indirect verification of the MSSM virtual particles in our sets at the ILC.

Since high luminosity is expected at the ILC experiments \cite{ilcd}, the quite small experimental error is expected. Correspondingly, therefore, the quite accurate calculations of the physical observables are required. As an explicit model, the Minimal Supersymmetric Standard Model (MSSM) is studied in this paper.
We have calculated cross sections and decay branching ratios at the 1-loop level using the $\tt GRACE/SUSY$-$\tt loop$ system \cite{wf,Kouda:2016ifp,jimbo,qcd14,grace2}. 
In this paper we report numerical results on the cross section of $e^-e^+\rightarrow Zh$\cite{jaque}.
Our results are consistent with those given in the previous work \cite{Cao2014rma} under the same setting of MSSM parameters. 
We have also calculated the cross section of  $e^-e^+\rightarrow \nu\bar{\nu}h$,  with the equivalent $\it W $-boson approximation (EWA) \cite{ewa2}. 
The complete SM 1-loop correction of $\sigma(e^-e^+\rightarrow Zh)$ and $\sigma(e^-e^+\rightarrow \nu\bar{\nu}h)$ have already been calculated using the {\tt GRACE} \cite{smnnh} and we reproduced the results for comparison.
We calculate the MSSM 1-loop correction to perform the indirect search of sparticles as they contribute to the cross sections through 1-loop diagrams. \\
　The parameter sets of the MSSM are chosen so as to meet the various experimental constraints. They include the anomalous magnetic moment of muon $g$-$2$ \cite{g-2exp,g-2}, the dark matter (DM) thermal relic density \cite{dmmssm1,dmmssm2,dmob}, $\it B$ meson rare decay branching ratio Br($b\rightarrow s\gamma$) \cite{btosg} and the higgs mass   \cite{higgsatlus,higgscms}. By the constraints the mass spectra of MSSM are quite limited.
For the selection of MSSM mass spectra we utilized the following program packages.
{\tt MicrOMEGAs} \cite{micromega} was used in the estimation of the DM thermal relic density  and {\tt SuSpect2} \cite{suspect2} was used for $g$-$2$ , Br($b\rightarrow s\gamma$)  and $m_h$.
 
% Since high luminosity is expected  at the ILC experiments \cite{ilcd},
%it is required to calculate the  physical observables with the accuracy better than the experimental data. Therefore we have calculated several cross sections and decay branching ratios at the 1-loop level with  $\tt GRACE/SUSY$-$\tt loop$ system \cite {grace2,jimbo}. In this paper we report numerical results on the cross section of $e^-e^+\rightarrow$ $Zh$ ($\sigma_{Zh}$)\cite{jaque}. Our results consistent with the one given in the previous work \cite{Cao2014rma} under same setting of MSSM parameters.
%We also calculeted the $e^-e^+\rightarrow \nu \bar \nu h$ ($\sigma_{\nu \bar \nu h}$) with the ``effective $W$ boson approximation"(EWA)\cite{ewa}. The complete SM 1-loop correction has already been calculated using $\tt GRACE$ \cite{smnnh}. For the selection of MSSM mass spectra we used next program packages. We used MicrOMEGAs \cite{micromega} in the estimation of the dark matter (DM) thermal relic abundance \cite{dmmssm1,dmmssm2,dmob}. 
%SuSpect2\cite{suspect2} was used
%for calculation of the muon $g$-$2$ anomalous magnetic moment \cite{g-2exp,g-2}($g$-$2$), the rare decay branching ratios of $B$ mesons \cite{btosg}($b\to s\gamma$), and the observed higgs masses \cite{higgsatlus,higgscms}.

\section{The selection of the MSSM parameter sets}
　In Table 1 we show experimental constraints we have
considered, and in Table 2 we show three MSSM parameter sets we have selected.
Detailed methods for the selection of sets have been explained in the previous work \cite{Kouda:2016ifp}. \\

%\hskip30pt $\cdots$
The first common settings in the three sets are \\
(A) $M_1=350$GeV,  $M_2=450$GeV, $\mu=1$TeV,  $\tan\beta=$50 and $m_{\tilde{\ell}} < 500$GeV ($\ell =e, \mu$). \\
They are necessary for sets to satisfy the constrains (1), (2) and (3) in Table.1. 
 As a consequence, the lightest sparticle (LSP) is almost Bino $\tilde{\chi}_1^0 \simeq \tilde{B}$ with mass about $350$GeV,  
$m_{\tilde{\chi}_2^0} \simeq m_{\tilde{\chi}_1^+} \simeq 450$GeV (almost Wino $\tilde{W}$) and 
$m_{\tilde{\chi}_{3,4}^0} \simeq m_{\tilde{\chi}_2^+} \simeq 1$TeV (almost higgsino $\tilde{H}$). 
The second common settings are \\
(B) a large mass splitting in the stau $\tilde{\tau}$ sector, 
$m_{{\tilde{\tau}}_1}\simeq	\emph{} 365$GeV ($\nge m_{\tilde{\chi}_1^0}$), 
$m_{{\tilde{\tau}}_2}\simeq 554$GeV and 
$m_{{\tilde{\nu}}_\tau}\simeq 466$GeV.  \\
They are necessary for sets to be satisfied the constraint (4),
because the co-annihilation occurring between $\tilde{\tau}_1$ and the LSP is required for meeting this constraint 
in the Bino LSP case. The difference among the three sets exists in the settings of masses  of the strongly interacting sparticles ($m_{\tilde{g}}$ ($=M_3$) and $m_{\tilde{q}}$), 
gluino and squarks. 
Considering the LHC bounds (5), we take 
 ($m_{\tilde{g}}$ and $m_{\tilde{q}}$) $\simeq$ ($2$TeV, $1.5$TeV),  ($5$TeV, $5$TeV) and ($10$TeV, $10$TeV) for 
 set 1, set 2 and set 3, respectively. 
For each set, the left-right mixing parameter $\theta_t$ and masses $m_{\tilde{t}_{1,2}}$ in the stop $\tilde{t}$ sector are 
tuned to satisfy the higgs mass constraint (6).
\clearpage
\begin{table}[h]
\caption{\label{tabone}The experimental constraints for MSSM parameters.}
\begin{center}
\begin{tabular}{*{8}{l}}                    
\hline
&Experimental bounds\\
\hline
\rm (1) muon $g$-$2$ anomalous magnetic moment \cite{g-2exp}&\rm$a_\mu^{\rm exp}-a_\mu^{\rm SM}=(2.88\pm0.63\pm0.49)\times 10^{-9}$ \\
\hline
\rm(2) $B$ meson rare decay branching ratio \cite{btosg} & $\rm Br($\it B$\rightarrow \chi_s\gamma) = (3.43\pm0.21\pm0.07)\times10^{-4}$\\
\hline
\rm (3) LHC direct search of chargino and neutralino \cite{atlasch}\cite{cmsch} & $m_{\tilde \chi^0_1} \nge 300$ GeV for $m_{\tilde\chi^\pm_1,\tilde\chi^0_2} = 500$ GeV\\
\hline
\rm (4) DM thermal relic density \cite{dmob} & $\it \Omega h^{\rm 2}=\rm 0.1198\pm0.0026$ \\
\hline
(5) \rm LHC direct search of gluino squark  \cite{atlasgl}\cite{cmsgl} & $m_{{\tilde q}}, m_{\tilde g}\nge 1.5 \rm TeV$\\
\hline
\rm (6) higgs mass \cite{higgsatlus}\cite{higgscms} &$ m_h(exp)=125.09\pm0.24$GeV\\
\hline
\end{tabular}
\end{center}
\end{table}

\begin{table}[h]
\caption{Masses and MSSM parameters for three sets (masses in unit of GeV)}
\label{table3}
\centering
\scalebox{1}{
\begin{tabular}{|c|c|c|c||c|c|c|c||c|c|c|c|}%%%The number of columns 
\hline
\multicolumn{4}{|c||}{set 1} & \multicolumn{4}{|c||}{set 2} & \multicolumn{4}{|c|}{set 3} \\
\hline
\hline
$\tilde\chi^+_1$ &  $\tilde\chi^+_2$ &  &  &  $\tilde\chi^+_1$ &   $\tilde\chi^+_2$ &   &   &  $\tilde\chi^+_1$ &   $\tilde\chi^+_2$ &   &   \\
\hline
456.2  & 1008 &  & &456.2  & 1008 &  &  &  456.2  & 1008 &  &\\
\hline
\hline
   $\tilde\chi^0_1$ &  $\tilde\chi^0_2$ &    $\tilde\chi^0_3$ & $\tilde\chi^0_4$& $\tilde\chi^0_1$ &  $\tilde\chi^0_2$ &    $\tilde\chi^0_3$ & $\tilde\chi^0_4$ & $\tilde\chi^0_1$ &  $\tilde\chi^0_2$ &    $\tilde\chi^0_3$ & $\tilde\chi^0_4$ \\
\hline
349.2 & 456.2 & 1003 & 1007 &    349.2 & 456.3 &  1003 & 1007 & 349.2 & 456.2 & 1003 & 1007 \\
\hline 
\hline
$\tilde \ell_1$ &  $\tilde \ell_2$ &  $\tilde\nu_\ell$ & & $\tilde \ell_1$ &  $\tilde \ell_2$  &  $\tilde\nu_\ell$ & &  $\tilde \ell_1$ &  $\tilde \ell_2$ &    $\tilde\nu_\ell$ &  \\
\hline
452.3 &471.9 &  465.7 & & 472.2 & 472.2 &  465.9 & & 472.2 & 471.9 &  465.8 & \\
\hline
\hline
$\tilde\tau_1$ &  $\tilde\tau_2$ &    $\tilde\nu_\tau$ &  & $\tilde\tau_1$ &  $\tilde\tau_2$ &    $\tilde\nu_\tau$ &  & $\tilde\tau_1$ &  $\tilde\tau_2$ &    $\tilde\nu_\tau$ &  \\
\hline
364.9 & 553.8 & 465.7 &  &  365.0 & 553.8 &  465.9 & & 364.9 & 554.5 &  465.8 & \\
\hline
\hline
$\tilde u_1$ & $\tilde u_2$ & $\tilde d_1$ & $\tilde d_2$ &$\tilde u_1$ & $\tilde u_2$ & $\tilde d_1$ & $\tilde d_2$ &$\tilde u_1$ & $\tilde u_2$ & $\tilde d_1$ & $\tilde d_2$ \\
\hline
1499 & 1500& 1500 & 1501& 5000 & 5000 & 5000 & 5000 & 10000 & 10000 & 10000 & 10000 \\
\hline
\hline
 $\tilde t_1 $ & $\tilde t_2$ & $\tilde b_1$ & $\tilde b_2$ & $\tilde t_1 $ & $\tilde t_2$ & $\tilde b_1$ & $\tilde b_2$ & $\tilde t_1 $ & $\tilde t_2$ & $\tilde b_1$ & $\tilde b_2$ \\
\hline
1413 & 1593 & 1461 & 1539& 4774 & 5220 & 4990 & 5010 & 9865 & 10130 & 9994 & 10010 \\
\hline
\hline
 $\theta_\tau$ &   $\theta_b$  &  $\theta_t$   &  & $\theta_\tau$ &   $\theta_b$  &  $\theta_t$   &  & $\theta_\tau$ &   $\theta_b$  &  $\theta_t$   & \\
\hline
0.7969  &   0.7913  &  0.7840 & &  0.8030 &  0.7912  &  0.7852  & & 0.8010  &   0.7912  &  0.7853  & \\
\hline
\hline
 $M_1$ &   $M_2$  &  $M_3$   & &  $M_1$ &   $M_2$  &  $M_3$   & &  $M_1$ &   $M_2$  &  $M_3$   & \\ 
\hline
350.0  &   450.0  &  2000   & & 350.0  &   450.0  &  5000   & & 350.0  &   450.0  &  10000   &  \\
\hline
\hline
\multicolumn{4}{|c||} {$\mu$=1000,~~~$\tan\beta$=50} & \multicolumn{4}{|c||} {$\mu$=1000,~~~$\tan\beta$=50} & \multicolumn{4}{|c|} {$\mu$=1000,~~~$\tan\beta$=50} \\
\hline
\end{tabular}
}
\end{table}
\clearpage
%\begin{table}[t]
%\caption{Main decay modes of the major MSSM particles.}%%%Table caption goes here
%\label{table4}
%\centering
%\scalebox{1.0}{
%\begin{tabular}{|c|c|c|c|c|c|c|c|}%%%The number of columns has to be defined here
%\hline
% &  $\tilde t_1 $ & $\tilde b_1$ & $\tilde g$ &  $\tilde\chi^+_1$ & $\tilde\chi^0_2$ & $\tilde\ell_{1,2}$   \\ %%%% Table body
%\hline
%  set 1 & $t \tilde\chi^0_1$, $t \tilde\chi^0_4$ & $b$$\tilde\chi^0_1$,  $t$$\tilde\chi^-_1$ & $t \tilde t_1 $, $b \tilde b_1 $  & $\tilde \tau_1 \nu_\tau$,$\tilde \chi^0_1 \it{W}$ & $\tau \tilde\tau_1$, $\tilde \chi^0_1 \it Z$ & $\ell$$\tilde\chi^0_1$ \\%%%% Table body
%\hline
% set 2 &  $b \tilde\chi^-_2$, $t \tilde\chi^0_4$  & $t$$\tilde\chi^-_2$, $b$$\tilde\chi^0_1$  & $t \tilde t_1 $, $b \tilde b_1 $  & $\tilde \tau_1 \nu_\tau$,$\tilde \chi^0_1 \it{W}$ & $\tau \tilde\tau_1$, $\tilde \chi^0_1 \it Z$ &$\ell$$\tilde\chi^0_1$ \\
%\hline
%set 3 & $b \tilde\chi^-_2$, $t \tilde\chi^0_4$ &$t$$\tilde\chi^-_2$, $b$$\tilde\chi^0_1$ & $t$$\tilde t_1 $, $b$$\tilde b_1 $ &  $\tilde \tau_1 \nu_\tau$,$\tilde \chi^0_1 \it{W}$ & $\tau \tilde\tau_1$, $\tilde \chi^0_1 \it Z$  & $\ell$$\tilde\chi^0_1$ \\
%\hline
%\end{tabular}
%}
%\end{table}
%\newpage
\section{Calculation scheme}
\subsection{The {\tt GRACE} system}
\,　There are more than twice as many different types of particles in the MSSM as those in the SM ; \\
therefore, there are various possible sparticle production processes in the collider experiments. 
A large number of Feynman diagrams appearing in each production process
requires tedious and lengthy calculations in evaluating the cross sections.
 Accurate theoretical prediction requires an automated system to manage  such large scale computations.
The $\tt GRACE$ system for the MSSM calculations\cite{grace1,grace2} has been developed by the KEK group (the Minami-tateya group) to meet the requirement. 
The $\tt GRACE$
system uses a renormalization prescription that imposes mass shell conditions on as many
particles as possible, while maintaining the gauge symmetry by setting the renormalization
conditions appropriately\cite{grace2}.
In the $\tt GRACE$ system for the SM, the usual 't Hooft-Feynman linear gauge 
 condition is generalized to a more general non-linear gauge (NLG) 
 that involves five extra parameters \cite{B, grace3}. 
  We extend it to the MSSM formalism by adding the SUSY interactions 
  with seven NLG parameters \cite{grace2, Baro}. We can check the consistency of the gauge symmetry by verifying the independence of the physical results from the NLG parameters. 
We ascertain that the results of the automatic calculation are reliable by carrying out the following checks:
\begin{itemize}
\item{Electroweak (ELWK) non-linear gauge invariance check (NLG check)}
\item{Cancellation check of ultraviolet divergence (UV check)} 
\item{Cancellation check of infrared divergence (IR check)}
\item{Check of soft photon cut-off energy independence ($k_c$ check)} 
\end{itemize}
 
Actually, the 1-loop differential cross sections (distributions) are separated into two parts, 
 \begin{equation}
 d\sigma_{\rm L\&S}^{\rm M}(k_c) \equiv d\sigma_{\rm virtual}^{\rm M}  + d\sigma_{\rm soft}(k_c),
 \end{equation}
where the suffix M indicates SM or MSSM, and each part is computed separately. The loop and the counter term contribution $d\sigma_{\rm virtual}^{\rm M}$ should be gauge invariant and the UV finite but IR divergent. 
We regularize the IR divergence by the fictitious photon mass $\lambda$, so both $d\sigma_{\rm virtual}^{\rm M}$ and the soft photon contribution 
$d\sigma_{\rm soft}$ are $\lambda$ dependent. 
The $\lambda$ dependence is canceled in $d\sigma_{\rm L\&S}^{\rm M}$.
Finally, the $k_c$ independent 1-loop physical cross sections can be obtained by

\begin{equation}
 d\sigma_{\rm 1loop}^{\rm M} \equiv d\sigma_{\rm tree} + d\sigma_{\rm L\&S}^{\rm M}(k_c)+ \int\int_{k_c}{ {\frac{d\sigma_{\rm hard}}{d\Omega dk}}d\Omega dk},
 \end{equation}
where, $k$ and $\Omega$ are the energy and the solid angle of the photon respectively. We selected the values of the MSSM parameters related to the higgs sector so that the tree level cross section are numerically identical between
 the MSSM and the SM. Therefore, the suffix M can be neglected in $d\sigma_{\rm tree}$ and $d\sigma_{\rm hard}$.
For the confirmation of the verifiability of 1-loop correction,
we defined following correction ratios,
\begin{equation}
\delta^{\rm M}_{\rm NLO}\equiv{\frac{d\sigma_{\rm 1loop}^{\rm M}-d\sigma_{\rm tree}}{d\sigma_{\rm tree}}}.
\end{equation}
%の比の数値結果を図3に示した。この結果から解るように、1-loop補正は
%cos$\theta$=0で、
%\begin{eqnarray}
%\delta_{\rm MSSM}^{\rm Zh}=8.5\%\nonumber
%\end{eqnarray}
%$E_h$=150で
%\begin{eqnarray}
%\delta_{\rm MSSM}^{\rm \nu\bar\nu h}=-20\%\nonumber
%\end{eqnarray}
%であり、統計誤差よりも十分大きいため、1-loop効果を検証する意味は十分にある。
For the estimation of effects of the MSSM virtual particles,
we defined following ratio \cite{hollik}, 
\begin{equation}
\delta_{\rm susy}\equiv
\delta_{\rm NLO}^{\rm MSSM}-\delta_{\rm NLO}^{\rm SM}=
{\frac{d\sigma_{\rm 1loop}^{\rm MSSM}-d\sigma_{\rm 1loop}^{\rm SM}}{d\sigma_{\rm tree}}}
={\frac{d\sigma^{\rm MSSM}_{\rm L\&S}-d\sigma_{\rm L\&S}^{\rm SM}}{d\sigma_{\rm tree}}}.
\end{equation}

For the process $e^-e^+ \to Z h$, we calculated differential cross sections 
\begin{equation}
d\sigma_{\rm tree, 1loop} = {\left(\frac{d\sigma}{d\cos\theta_Z}\right)_{\rm tree, 1loop}} \label{eezh},  
\end{equation}
where $\theta_Z$ is the scattering angle of the $Z$-boson. 

\subsection{Equivalent {\it W}-boson Approximation (EWA)}
For the process $e^-e^+\to\nu\bar\nu h$, we should calculate the cross section,
\begin{eqnarray}
\sigma(s) \equiv \sum_{\ell=e,\mu,\tau}\sigma(e^-e^+ \to \nu_\ell \bar{\nu}_\ell h).  \label{sg}
\end{eqnarray}
However, the estimation of the MSSM full 1-loop correction for (\ref{sg}) is difficult even using the $\tt GRACE$ system
because the number of Feynman diagrams of the process $e^-e^+\to \nu\bar\nu h $ becomes about $10^4$.\\
　In this paper, therefore, we use the EWA formulae \cite{ewa2},
\begin{eqnarray}
\sigma'(s) \equiv \sum_{n=-1}^{+1}\int_{x_{min}}^1 f_n(x) \hat{\sigma}_n(\hat{s})dx, \label{sgd} \end{eqnarray}
where $\hat{s}\equiv x s$ and $\hat{\sigma}_n(\hat{s})$ denotes the center of mass energy 
squared and cross section of the sub-process $e^- W_n^+ \to \nu_e h$. Then we set $x_{min} = (m_e+m_W)^2/s$. 
$f_n(x)$ are the energy distribution functions for $W_n$-boson with helicity $n=-1,0,+1$, 
\begin{eqnarray}
&&f_0(x) =(g_L^2+g_R^2)\left({\frac{x}{16\pi^2}}\right)\left[
     {\frac{2(1-x)\zeta}{\omega^2 x}}-{\frac{2\Delta (2-\omega)}{\omega^3}}\ln\left(\frac{x}{\Delta'}\right)
     \right], \\
&&f_{+1}(x) = g_L^2 h_2+g_R^2 h_1,  \\
&&f_{-1}(x) = g_L^2 h_1+g_R^2 h_2, 
\end{eqnarray}
where
\begin{eqnarray}
&& h_1 = {\frac{x}{16\pi^2}}\left[
              {\frac{-(1-x)(2-\omega)}{\omega^2}}
             +{\frac{(1-\omega)(\zeta-\omega^2)}{\omega^3}}\ln\left({\frac{1}{\Delta'}}\right)
             -{\frac{\zeta-2x\omega}{\omega^3}}\ln\left({\frac{1}{x}}\right)
             \right] ,\\     
&& h_2 = {\frac{x}{16\pi^2}}\left[
              {\frac{-(1-x)(2-\omega)}{\omega^2 (1-\omega)}}
              +{\frac{\zeta}{\omega^3}}\ln\left({\frac{x}{\Delta'}}\right)
              \right],
\end{eqnarray}
$\omega = x-\Delta$, $\zeta=x+\Delta$, $\Delta=m_W^2/s$, $\Delta'=\Delta/(1-\omega)$. 
For the electron and $W$-boson coupling,  $g_L=e/(\sqrt{2}\sin\theta_W)$ and $g_R=0$.
\begin{figure}[h]
\begin{center}
\includegraphics[width=.4\textwidth, angle=0]{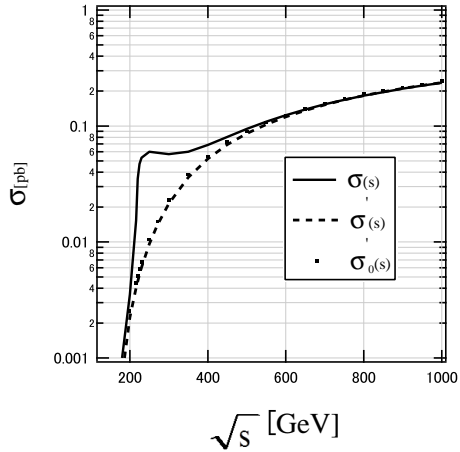}
\end{center}
\vspace{-20pt}
\caption{$\sqrt{s}$ dependence of tree level total cross sections of $e^-e^+\to \nu\bar{\nu} h$. 
The solid line and the dashed line and the dotted points correspond to $\sigma$, $\sigma'$ and $\sigma'_0$, respectively. }
\label{fig1}
\end{figure}
Moreover, the EWA cross section (\ref{sgd}) is numerically indistinguishable from
\begin{eqnarray}
\sigma'_{0}(s) \equiv \int_{x_{min}}^1 f_0(x)  \hat{\sigma}_0(\hat{s})dx, \label{sgd0}
\end{eqnarray}
in the energy range considered here, because the contribution of $W_{\pm}$ is almost negligible
as shown in Figures.\\
　In Figure 1, $\sqrt{s}$ dependence of the tree level cross sections (\ref{sg}), (\ref{sgd}) and (\ref{sgd0}) are shown. 
The cross section of $e^-e^+ \to Z h\to\nu\bar\nu h$ is apparently large in the region  $\sqrt{s} \lsim 400$GeV.
We calculated the differential cross section %(\ref{eezh})
 for $e^-e^+ \to Z h$ at $\sqrt{s}=250$GeV, 
where its total cross section has almost the largest value. 
On the other hand, the $W$-fusion contribution is dominant at  $\sqrt{s} \gsim 500$GeV and the EWA becomes good approximation.
For example, $(\sigma-\sigma')/\sigma \simeq$ ($-8.4$, $+0.75$)\% and 
$(\sigma-\sigma'_0)/\sigma \simeq$ ($-4.7$, $+3.2$)\% at $\sqrt{s}=$ ($500$, $1000$) GeV. 
We use $\sigma'_0$ for the calculation of cross sections at $\sqrt{s}=500$GeV. We calculated the differential cross section,
\begin{equation}
d\sigma_{\rm tree, 1loop} = \left({\frac{d\sigma'_0}{dE_h}}\right)_{\rm tree, 1loop},  
\end{equation}
where $E_h$ is the energy of higgs particle. 
It is calculated from %(\ref{sgd0})
(\ref{sgd0}) using  relations 

\begin{eqnarray}
&&{\frac{d\sigma'_0}{dE_h}}= {\frac{2 x}{\sqrt{E_h^2-m_h^2}}}{\frac{d\sigma'_0}{dx}}, \\
&& x={\frac{1}{s}}\left( E_h + \sqrt{E_h^2-m_h^2} \right)^2. 
\end{eqnarray}

\begin{figure}[h]
\begin{center}
\includegraphics[width=.40\textwidth, angle=0]{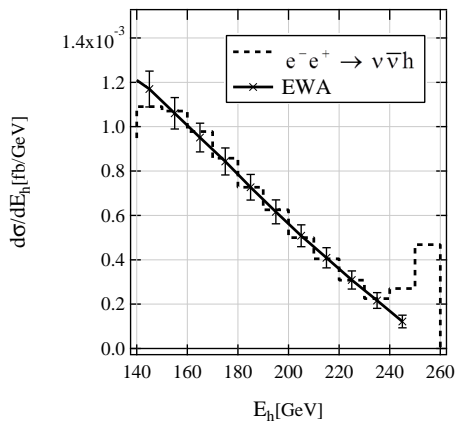}
\end{center}
\vspace{-20pt}
\caption{%The dependence on $\sqrt s$ of $e^-e^+\to Zh$ and $e^-e^+\to \nu\bar\nu h$. In the left figure, The thick solid line stands for the sum of $\rm \nu_e\bar\nu_e $$h$, $\rm \nu_\mu\bar\nu_\mu $$h$ and $\rm \nu_\tau\bar\nu_\tau $$h$ contributions. The dash-dotted line corresponds to $\sigma(e^-e^+\to Zh)\times \rm {Br}(\it Z\to\nu\bar\nu)$, In right figure, the solid curve corresponds to EWA}
The higgs energy distribution at tree level of $e^-e^+\to \nu\bar\nu h$ (dotted), and that with EWA(solid).  Here we assume $L$= 500 $\rm fb^{-1}$.}
\label{fig7}
\end{figure}
In Figure 2, we show the tree level energy distribution of the higgs $d\sigma/dE_h$ from (6) and $d\sigma^{'}_{0}/dE_h$ in (15)  at $\sqrt s$=500 GeV. The statistical errors were calculated assuming $L$=500$\rm fb^{-1}$.
The difference is within the error range in the region $E_h$=150$\sim$230 $\rm GeV$.
We could figure out that the EWA reproduce the $e^-e^+\to\nu\bar\nu h$ cross section in this energy region.
The peak from $\it Zh$ production around $E_h$=250 GeV cannot be reproduced by the EWA.
\section
{The numerical results}
%　We show the numerical results in following order.((1) is necessary to confirm that constructed our methods including EWA work well, and the 1-loop correction itself is sizable enough. (2) is the purpose of this study)\\

%\noindent(1) The confirmation of significance for the 1-loop correction of SM  (Figure 3) \\
%(2) The verifiability for the MSSM 1-loop effects (Figures 4,5)\\

\subsection{The SM contribution}
 Table 3 shows the SM parameters which we used in the calculations. All the following numerical results (include the MSSM cross section) are computed in these parameters.
Figure 3 shows the $\delta^{\rm SM}_{\rm NLO}$ in (3) for $e^-e^+\to Zh$ at $\sqrt s$=250 GeV (left) and $e^-e^+\to \nu\bar\nu h$ used EWA at $\sqrt{s}$=500GeV (right).
The plotted errors were calculated assuming $L$=250 $\rm{fb^{-1}}$ and 500 $\rm{fb^{-1}}$ for $e^-e^+\to Zh$ and $e^-e^+\to \nu\bar\nu h$ respectively. In advance, we should confirm that the 1-loop correction does not buried in this error.
In the left figure, the $\delta^{\rm SM}_{\rm NLO}$ is estimated to be $-(6\sim9)$\% in entire region and it is larger than the error.
Similarly, in the right figure,   $\delta^{\rm SM}_{\rm NLO}$ is estimated to be $-4.5\sim+1.5$\% in entire region and this value is larger than the error.
From the results we figure out that the 1-loop correction is necessary for reliable theoretical predictions in both processes. 
%次の図4に補正比$\delta_{NLO}$を示した。
%左の図が$e^+e^-\to Zh$右の図が$e^+e^-\to \nu\bar\nu h$
%であり誤差はL=500$\rm fb^-1$を仮定したときのtreeのイベント数を元にした統計誤差である。
%この誤差は非常に素朴なものであるが、この誤差に1-loop効果が埋もれないことをまず
%はじめに確かめる必要があるだろう。
%左の図では全域で$-5\sim-10$\%であり、全域にわたり、誤差を上回っていることが解る。
%よって、このプロセスの1-loop効果を検証する意味があることが解る。
%右の図では、断面積の小さな領域に近づくに従って誤差が大きくなっている
%ので、高エネルギー側では1-loop効果は誤差に埋もれてしまう可能性もあるが、
%higgsのエネルギーが220GeV程度までであれば1-loop効果を検証する意味があることが解る。
\begin{table}[h]
\caption{\label{tabone}The SM parameters used in this report.}
\begin{center}
\begin{tabular}{|c|c||c|c|}                      
\hline
\rm {\it u}-quark mass &$2.0\times10^{-3}\rm GeV$ &{\it d}-quark mass & $5.0\times10^{-3}\rm GeV$ \\ 
\hline
\rm {\it c}-quark mass &$1.0\times10^{-3}\rm GeV$ & {\it s}-quark mass &$100.0\times10^{-3}\rm GeV$ \\ 
\hline
\rm {\it t}-quark mass &173.5\rm GeV &{\it b}-quark mass& $4.75\times10^{-3}\rm GeV$\\ 
\hline
\rm {\it W} boson mass& 80.4256\rm GeV & {\it Z} boson mass & 91.187\rm GeV\\  
\hline
\rm higgs mass & 125.1GeV&QED fine structure constant &1/137.036 \\  
\hline
\end{tabular}
\end{center}
\end{table}

\begin{figure}[h]
\begin{center}
\includegraphics[width=.75\textwidth, angle=0]{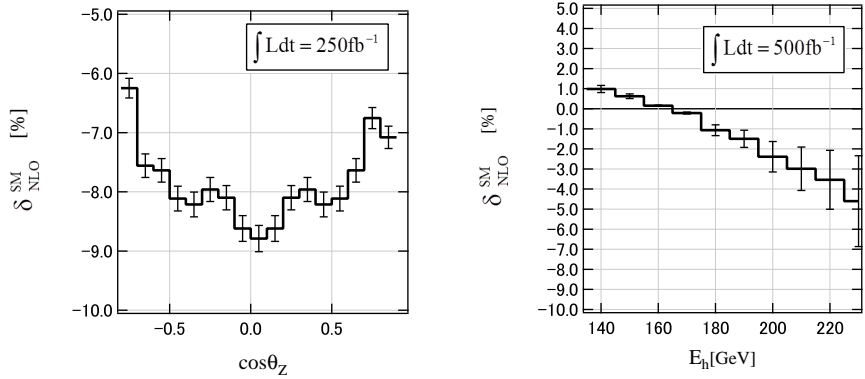}
\end{center}
\vspace{-20pt}
\caption{The correction ratio $\delta^{\rm SM}_{\rm NLO}$.  $e^-e^+\to Zh$ with ($\sqrt s, L$)= (250 GeV, 250 ${\rm fb}^{-1}$) (left), $e^-e^+\to \nu\bar\nu h$ with ($\sqrt s, L$)= (500 GeV, 500 $\rm fb^{-1}$) (right).}
\label{fig10}  
\end{figure}

\subsection{The MSSM contribution}
%We show the angular distribution of $\it Z$ in $e^-e^+ \rightarrow Zh$ in Figure 3 left figure.
%For  $\sqrt{s}$ = 250 GeV, the 1-loop correction at cos$\theta$ = 0 is estimated to be\\
%\begin{eqnarray}
%\delta_{\rm SM} \equiv{\frac{(d\sigma^{\rm SM}_{\rm 1loop}/{d\cos\theta})-(d\sigma_{\rm tree}/{d\cos\theta})}{(d\sigma_{\rm tree}/{d\cos\theta})}}= -10\%\nonumber 
%\end{eqnarray}  \\and 
%\begin{eqnarray}
%\delta_{\rm MSSM} \equiv{\frac{(d\sigma^{\rm MSSM}_{\rm 1loop}/{d\cos\theta})-(d\sigma_{\rm tree}/{d\cos\theta})}{(d\sigma_{\rm tree}/{d\cos\theta})}}=-8.7\%. \nonumber
%\end{eqnarray}
%The magnitude fills following relation,
%\begin{eqnarray}
%\rm SM\nle set1<set2<set3<<tree.
%\end{eqnarray}
　Figure 4 shows the 1-loop corrected angular distribution of $\it Z$, $d\sigma/dcos\theta_{Z}$ (left) and the ratio $\delta_{\rm susy}$ in (4) (right) for the process $e^-e^+\to Zh$.
From the left figure, we find that the both the SM and the MSSM 1-loop contributions are negative and they are satisfied the relations,
\begin{eqnarray}
\frac{d\sigma^{\rm{SM}}_{\rm 1loop}}{d\rm{cos}\theta_Z} \nle \frac{d\sigma^{\rm set 1}_{\rm 1loop}}{d\rm{cos}\theta_Z}\nle \frac{d\sigma^{\rm set2}_{\rm 1loop}}{d\rm{cos}\theta_Z}\nle \frac{d\sigma^{\rm set3}_{\rm 1loop}}{d\rm{cos}\theta_Z}<\frac{d\sigma_{\rm tree}}{d\rm{cos}\theta_Z}.
\end{eqnarray}
 In the right figure,  
the $\delta_{\rm susy}$ are estimated to be 1.17$\sim$1.25\%
at $\rm cos\theta_{\it Z}$ = 0.
In the entire region these ratios are larger than the statistical error assuming planned luminosities at the ILC.  
It means that the 1-loop contribution of the MSSM could be measured at  $\sqrt s$ =250GeV.
Also, the set 1 will be possibly distinguished from set 2 and set 3.\\
%squarkの質量が大きいsetほどtreeに対する1-loop補正値は小さい。
%We define the ratio of the differential cross sections \cite{hollik},
%We show  $\rm \delta_{susy}$ in right figure.  $\rm \delta_{susy}$ is 1$\sim$2\% in the entire region and is larger than the statistical error assuming actually planned luminosities at the ILC.   It means that the 1-loop contribution of the MSSM could be measured at  $\sqrt s$ =250 GeV.
%In addition, the difference between set2 and set3 is also larger than the statistical error at $\sqrt s$=250 GeV.
\begin{figure}[h]
\begin{center}
\includegraphics[width=.75\textwidth, angle=0]{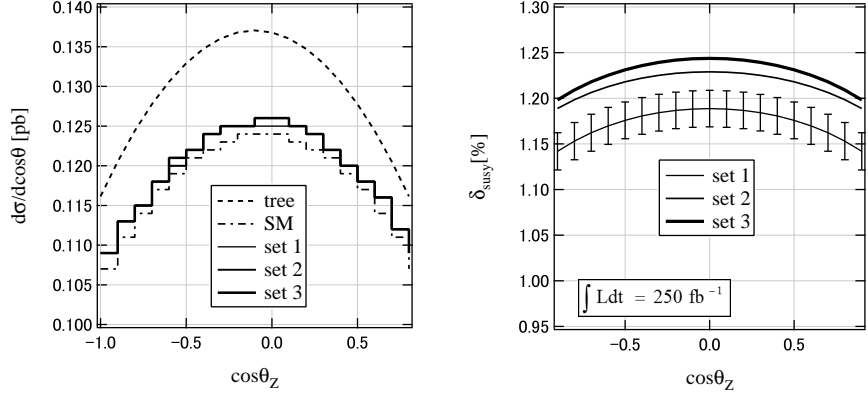}
\end{center}
\vspace{-20pt}
\caption{%The angular distribution of the cross section and $\delta_{\rm susy}$ at $\sqrt s$ = 250 GeV. The error bar in the right figure means statistical error which is assumed $L=250 \rm fb^{-1}$.}
The angular dependence  $d\sigma/d\rm cos\theta_Z$ (left), $\delta_{\rm susy}$(right) at $\sqrt s$ = 250 GeV in $e^-e^+ \rightarrow Zh$. The error bars mean the statistical error assuming  $L$ =  250 $\rm fb^{-1}$.} 
\label{fig10}
\end{figure}
%$\rm \delta_{susy}$ is 1.12$\sim$1.25\% in the entire region and is larger than the statistical error assuming actually planned luminosities %at the ILC.   It means that the 1-loop contribution of the MSSM could be measured.
%In addition, the difference between set2 and set3 is also statistically significant.
　Figure 5 shows the 1-loop corrected energy distribution of higgs (left) and $\delta_{\rm susy}$(right) in $e^-e^+\to \nu\bar\nu h$.
The left (right) figure shows differential cross section $d\sigma/d{E_h}$($\delta_{\rm susy}$).
In the left figure the SM and MSSM 1-loop contribution are indistinguishable at a glance.
However,
$\delta_{\rm susy}$ are estimated to be 1.5$\sim$1.7\%
and they are larger than the statistical error assuming 500 $\rm fb^{-1}$ in the entire region.  
It means that the 1-loop contribution of the MSSM could be measured at  $\sqrt s$ =500 GeV. 
Three proposed sets can not be distinguished each other at least assuming the planned luminosities at the ILC. The precision measurements of the higgs production processes at the ILC will bring us important information on the heavy sparticles through the virtual 1-loop effects.
%In the Next, let us see the $E_h$ distribution and the correction ratio $\rm \delta_{susy}$ in $e^-e^+\to \nu_e\bar\nu_e h$ which are shown in Figure 4. 
%Where $E_h$ is the higgs energy in single higgs productions.
%The left figure shows $\rm d\sigma/dE_h$, and the magnitude fill following relation again,
%\begin{eqnarray}
%\rm SM\nle set1<set2<set3<<tree.
%\end{eqnarray}
%$\rm \delta_{susy}$ of set1 and set2 (set3) is 0.8$\sim$1.0(2.2$\sim$3.1)\% in the entire region and is larger than the statistical error assuming actually planned luminosities at the ILC.   It means that the 1-loop contribution of the MSSM could be measured.
%In addition, the  set3 is statistically significant for other sets.
%It is larger than the statistical error assuming $L$ = 500 $\rm fb^{-1}$.
%$\sqrt s$=500GeVでの単独生成の数値結果を図4に示す。
\begin{figure}[h]
\begin{center}
\includegraphics[width=.77\textwidth, angle=0]{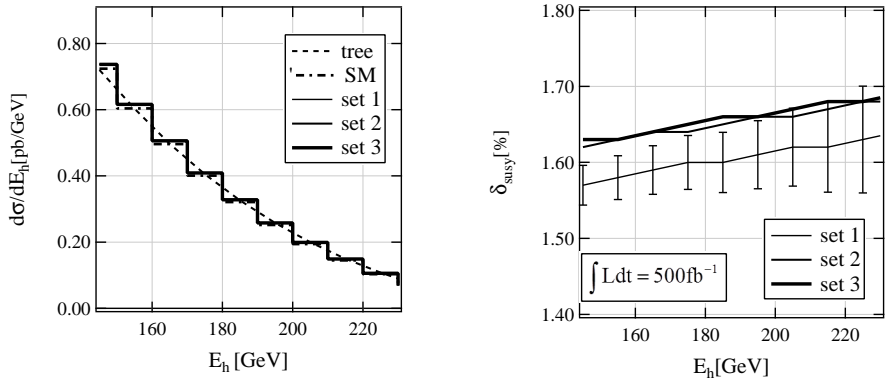}
\end{center}
\vspace{-20pt}
\caption{The energy dependence $d\sigma/dE_h$(left), $\delta_{\rm susy}$(right) of higgs at $\sqrt s$= 500 GeV in $e^- e^+ \rightarrow \nu \bar \nu h$. The error bars mean statistical error assuming $L$= 500 $\rm fb^{-1}$.}
\label{fig11}
\end{figure}

\section{Summary and conclusions}
　We investigated the indirect effects of the MSSM at the ILC.
We focused on the center of mass energies of 250 GeV and 500 GeV in the processes $e^-e^+\to Zh$ and $e^-e^+\to \nu\bar\nu h$, respectively. We selected the mass spectra of the MSSM which are consistent with the observed mass of higgs,  
the thermal relic density of the dark matter, the low energy experiments and the LHC bounds of sparticles.
The parameter sets proposed in our calculations have the squarks and the gluino with masses of 1.5, 5.0, 10.0 TeV and the sleptons and gauginos 
with masses less than 0.5 TeV. With using our developed $\tt GRACE$ system, we calculated the 1-loop corrected cross sections of the processes $e^-e^+\to Zh$ and $e^-e^+\to \nu\bar\nu h$ at the ILC. For the analysis of latter process, we adopted the equivalent $\it W$-boson approximation.
\\
　We confirmed that both the SM and the MSSM 1-loop correction is necessary for
the accurate theoretical predictions at the ILC. We found that the 1-loop effects of MSSM are verifiable both at 250 GeV and 500 GeV. Moreover the difference between set 1 and set 2 and 3 will be possibly observed with $e^-e^+\to Zh$ at $\sqrt s$=250GeV.
\vspace{5pt}


\begin{thebibliography}{10}

\bibitem{martin}
S.~P. Martin, Adv. Ser. Direct. High Energy Phys. {\bf 18}, 1 (1998),
  {{hep-ph/9709356v7 (2016)}}.

\bibitem{atlastop}
The ATLAS collaboration, JHEP., {\bf 1712}, 085 (2017). 
\bibitem{cmsstop}
The CMS collabolation, Phys. Rev. D, {\bf 97}, 032009 (2018).


\bibitem{ilcd}
H.~Baer, T.~Barklow, K.~Fujii, Y.~Gao, A.~Hoang, S.~Kanemura, J.~List, H.~E.
 Logan, A.~Nomerotski, M.~Perelstein, et~al. (2013),  {{arXiv:1306.6352}}.

\bibitem{wf}

Y.~Kouda, T.~Kon, Y.~Kurihara, T.~Ishikawa, M.~Jimbo, K.~Kato,. and M.~Kuroda, Journal of Physics Conference Series., {\bf 920-1}, 012010 (2017).

\bibitem{Kouda:2016ifp}
Y.~Kouda, T.~Kon, Y.~Kurihara, T.~Ishikawa, M.~Jimbo, K.~Kato,. and M.~Kuroda,  Prog. Theor. Exp. Phys., {\bf 2017-5},  053B02 (2017).

\bibitem{jimbo}

 M. Jimbo, T. Kon, Y.Kouda,. M Ichikawa, Y Kurihara, T Ishikawa, K Kato, and M. Kuroda, (2017), {eConf C16-12-05.4 [arXiv:1703.07671]}.

\bibitem{qcd14}
M. Kuroda, T. Kon, Y. Kouda, T. Ishikawa, M. Jimbo, K. Kato and Y. Kurihara. Nuclear and Particle Physics Proceedings., {\bf 258-259},  264 (2015).

\bibitem{grace2}
J.~Fujimoto, T.~Ishikawa, Y.~Kurihara, M.~Jimbo, T.~Kon, and M.~Kuroda, Phys.
 Rev. D, {\bf75}, 113002 (2007).





\bibitem{jaque}

J. Yan  S. Watanuki, K. Fujii, A. Ishikawa, D. Jeans, J. Strube, J.Tian, H. Yamamoto,
 Phys. Rev. D, {\bf94,} 113002 (2016).

\bibitem{Cao2014rma}
J.~Cao, C.~Han, J.~Ren, L.~Wu, J.~Yang, M.~Jin and Y.~ Zhang, Chin. Phys. C, {\bf40}, 113104 (2016).


\bibitem{ewa2} 
P.~W.~Johnson, F.~I.~Olness, and Wu-Ki Tung, Phys. Rev. D, {\bf36}, 291 (1987).

\bibitem{smnnh}
G. B\'{e}langer, F. Boudjema, J. Fujimoto, T. Ishikawa , T. Kaneko, K. Kato, and Y. Shimizu, Phys. Lett. B, {\bf 559}, 252 (2003). 

\bibitem{g-2exp}
A.~Hoecker and W.~J. Marciano, The Muon Anomalous Magnetic Moment, in Particle
  Data Group, Chin. Phys. C {\bf 38}, 090001 (2014); p.649 (updated August 2013), references therein.
\bibitem{g-2}
G.~C. Cho, K.~Hagiwara, Y.~Matsumoto, and D.~Nomura, JHEP., {\bf11}, 068
  (2011).

\bibitem{dmmssm1}
A.~Ibarra, A.~Pierce, N.~R. Shah, and S.~Vogl, Phys. Rev. D, {\bf 91}, 095018
  (2015).

\bibitem{dmmssm2}
J.~Ellis, K.~A. Olive, and J.~Zheng, Eur. Phys. J. C, {\bf74}, 2947 (2014).

\bibitem{dmob}
K.~A. Olive, PoS, {\bf PLANCK2015}, 093 (2015).

\bibitem{btosg}
T.~Becher and M.~Neubert, Phys. Rev. Lett, {\bf 98}, 022003 (2007).

\bibitem{higgsatlus}
The ATLAS Collaboration, Phys. Lett. B, {\bf716}, 1 (2012).

\bibitem{higgscms}
The CMS Collaboration, Phys. Lett. B, {\bf716}, 30 (2012).

\bibitem{micromega}
G.~B\'{e}langer, F.~Boudjema, A.~Pukhov, and A.~Semenov, Comput. Phys. Commun.,
  {\bf 149}, 103 (2002).

\bibitem{suspect2}
A.~Djouadi, J.-L. Kneur, and G.~Moultaka, Comput. Phys. Commun., {\bf176}, 42 (2007).


\bibitem{atlasch}
The ATLUS collaboration, Eur. Phys. J. C, {\bf78},  154 (2018).

\bibitem{cmsch}

The CMS collaboration, JHEP., {\bf 1803}, 160 (2018).

\bibitem{atlasgl}

The ATLAS collaboration Phys. Rev. D, {\bf 96}, 112010 (2017).

\bibitem{cmsgl}

The CMS collaboration, JHEP., {\bf1712}, 142 (2017).

\bibitem{grace1}
J.~Fujimoto, T.~Ishikawa, M.~Jimbo, T.~Kaneko, K.~Kato, S.~Kawabata, T.~Kon,
 M.~Kuroda, Y.~Kurihara, Y.~Shimizu, and H.~Tanaka, Comput. Phys. Commun.,
  {\bf 153}, 106 (2003).


%\bibitem{ag2}
%ATLAS Collaboration, JHEP, {\bf 10}, 134 (2015).

%\bibitem{nthg}

%Hall, Lawrence J. and Pinner, David and Ruderman, Joshua T., JHEP {\bf04}, 131(2012)

%\bibitem{kitano}

%Kitano, Ryuichiro and Nomura, Yasunori, Phys. Lett. {\bf B631}, 58-67 (2005)

%\bibitem{CERN-TH.4825/87}

%Barbieri, Riccardo and Giudice, G. F.,Nucl. Phys. {\bf B306}, 63-76, CERN-TH-4825/87(1987)
\bibitem{B}
F.~Boudjema and E.~Chopin, Z. Phys., {\bf 73}, 85 (1997).

\bibitem{grace3}
G.~B\'{e}langer, F.~Boudjema, J.~Fujimoto, T.~Ishikawa, T.~Kaneko, K.~Kato, and
  Y.~Shimizu, Nucl. Phys. Proc. Suppl., {\bf 116}, 353 (2003).


\bibitem{Baro}
N.~Baro and F.~Boudjema, Phys. Rev. D, {\bf 80}, 076010 (2009).
 


\bibitem{hollik}
W.~Hollik and C.~Schappacher, Nucl. Phys.  B, {\bf545}, 98 (1999).

%\bibitem{kane}

%S.~kanemura, T.~Shindou, K.~Yagu., Phys. Lett., {\bf B699} 258-263 (2011).












%\bibitem{hamaguchi1}
%M.~Endo, K.~Hamaguchi, S.~Iwamoto, T.~Kitahara, and T.~Moroi, Phys. Lett., {\bf
%  B728}, 274 (2014).

%\bibitem{wang}
%B.~P. Padley, K.~Sinha, and K.~Wang, Phys. Rev. D, {\bf 92}, 055025 (2015).

%\bibitem{hamaguchi2}
%K.~Hamaguchi and K.~Ishikawa, Phys. Rev. D, {\bf 93}, 055009 (2016).



%\bibitem{susyhit}
%A.~Djouadi, M.~M. Muhlleitner, and M.~Spira, Acta Phys. Polon., {\bf B38}, 635
%  (2007).




%\bibitem{grace1}
%J.~Fujimoto, T.~Ishikawa, M.~Jimbo, T.~Kaneko, K.~Kato, S.~Kawabata, T.~Kon,
%  M.~Kuroda, Y.~Kurihara, Y.~Shimizu, and H.~Tanaka, Comput. Phys. Commun.,
%  {\bf 153}, 106 (2003).
%\bibitem{jimbo}

% M. Jimbo, T. Kon,  Y.Kouda,.  M Ichikawa, Y Kurihara, T Ishikawa, K Kato,  and M. Kuroda, (2017) {arxiv:1703.07671 [hep-ph]}

%\bibitem{grace2}
%J.~Fujimoto, T.~Ishikawa, Y.~Kurihara, M.~Jimbo, T.~Kon, and M.~Kuroda, Phys.
%  Rev. D, {\bf 75}, 113002 (2007).


%\bibitem{p}
%Planck Collaboration (2015),  {{arXiv:1502.01582}}.



%\bibitem{glsq}
%ATLAS Collaboration, JHEP, {\bf 09}, 176 (2014).

%\bibitem{neutralinochargino}
%ATLAS Collaboration, JHEP, {\bf 04}, 169 (2014).

%\bibitem{PhysRevD.36.724}

%K. Hikasa and M. Kobayashi,  Phys. Rev. D, {\bf 36}, 724--732 (1987).

%\bibitem{directstop}
%ATLAS Collaboration, ATLAS-CONF-2016-077, (2016).

%\bibitem{cms}
%The~CMS Collaboration, Eur. Phys. J. C, {\bf 73}, 2677 (2013).

%\bibitem{sbottom}
%ATLAS Collaboration, JHEP, {\bf 10}, 189 (2013).







\end{thebibliography}
\end{document}